\documentclass{iau}
\usepackage{graphicx}

\title[] %% give here short title %%
{On the relation of accretion rate and spin induced jet power in
low luminosity AGN}

\author[Xiang Liu, Zhen Zhang \& Zhenhua Han]   %% give here short author list %%
{Xiang Liu$^{1,2}$
\and Zhen Zhang$^{1}$
 \and Zhenhua Han$^{1,3}$}
\affiliation{$^1$Xinjiang Astronomical Observatory of CAS, 150 Science 1-Street, Urumqi 830011, China, email: {\tt liux@xao.ac.cn} \\[\affilskip]
$^2$Key Laboratory of Radio Astronomy, CAS, Nanjing 210008, China\\
$^3$Graduate University of Chinese Academy of Sciences,
Beijing 100049, China \\ %Box 515, SE-75120 Uppsala, Sweden
}

\pubyear{2015}
\volume{313}  %% insert here IAU Symposium No.
\pagerange{119--120}
\jname{Extragalactic Jets from every angle}
\editors{F. Massaro, C. C. Cheung, E. Lopez, A. Siemiginowska, eds.}
\begin{document}

\maketitle

\begin{abstract}
From \cite[Liu \& Han (2014)]{liu14}, the accretion-dominated jet
power has a linear proportionality with the accretion rate,
whereas the power law index is $\leq$0.5 at lower accretion rate.
Attributing the jet power in low accretion rate AGN to the black
hole spin, it implies that the jet power has a flatter spectrum
than the accretion-dominated jet versus the accretion rate. The
black hole must be spinning rapidly for producing such jet power
efficiently, and this may allow us to find high-spin black holes
in the radio-loud low-luminosity AGN.

\end{abstract}

\firstsection % if your document starts with a section,
%\section{Introduction}
\section{BZ-jet power may have a flatter spectrum versus accretion rate}

Radio jet of AGN could be produced from either the disk accretion
or the black hole (BH) spin, or even both. The accretion produced
jet power could be linearly proportional to the product of the Eddington ratio and BH mass, 
as found in radio loud quasars \cite[(Liu
\& Han 2014)]{liu14}. Whereas, in relative low luminosity AGN,
e.g. in spiral galaxies, the index of the power law relation between
jet power and accretion rate is significantly less than unity,
implying that besides the accretion, other mechanisms must be at
work in the production of jet, e.g. the BH spin. In this
paper, we investigate further in this aspect.

The BH spin can produce jet as suggested by \cite[Steiner et al.
(2013)]{ste13}, which principally predicted by \cite[Blandford \&
Znajek (1977)]{bla77}, BZ-jet hereafter. The BZ-jet depends on the
magnetic flux ($\Phi$) threading on the BH ergosphere and the spin
($a_{*}$) of BH, and can be expressed as the function of $a_{*}$
and relative magnetic flux $\phi=\Phi/\Phi_{max}$. The
maximum magnetic flux $\Phi_{max}$ has a root-squared relation
with the disk accretion rate \cite[(Tchekhovskoy et al.
2011)]{tch11} and see \cite[Yuan \& Narayan (2014)]{yuan14} for a
review, but it is not clear how the magnetic flux $\phi$ relates
to the accretion rate. We think that a minimum accretion rate is
required for the BH spin to produce a jet, since the magnetic flux
(outgoing Poynting flux) needs plasma from the accretion to
the BH horizon. There may also exist a critical accretion rate for
the maximum magnetic flux $\Phi_{max}$ as noted by
\cite[Tchekhovskoy et al. (2011)]{tch11} and \cite[Yuan \& Narayan
(2014)]{yuan14} for the BZ-jet. Exceeding the critical accretion
rate (or the critical Eddington ratio $\lambda_{c}$ for similar BH
masses), the BZ-jet could be suppressed, and the accretion disk
would dominate the behavior of total jet power when the
accretion rate is greater than the $\lambda_{c}$.

There is a false linear proportional relation between the BZ-jet
power and the mass accretion rate (see \cite[Yuan \& Narayan
2014]{yuan14}), because the $\Phi$ in that formula is also a
function of the accretion rate. The relation between the BZ-jet
power and the mass accretion rate should be a non-linear form
which we want to know. To consider both the accretion and the
BH spin contribute to jet power, the total jet power should be the
sum of the two but not the multiplying of two contributions.

\begin{figure}
\centering
    \includegraphics[width=7cm]{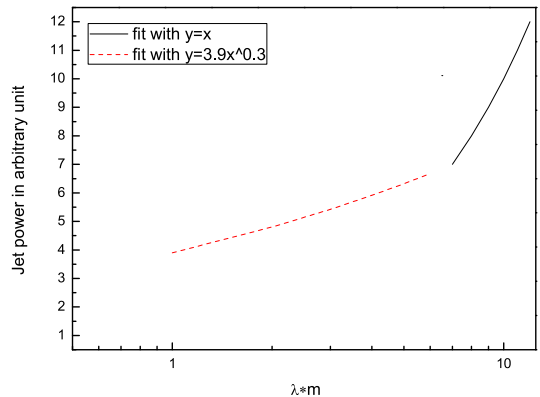}
    \caption{Jet power vs. the product of the Eddington ratio ($\lambda$)
    and BH mass in solar mass unit ($m$), for lower accretion AGN (dash line)
    and high accretion AGN (solid line), respectively.  }
     \label{fig1}
  \end{figure}

As the first approximation, we assume that the jet power is
dominated by the accretion at high accretion rate (e.g.
$\lambda> 0.1$, or $\lambda m>10^7$ for $10^{8}$ solar mass BH), and the jet
power is dominated by the BZ-jet at lower accretion rate ($\lambda< 0.1$, or $\lambda m<10^7$), as noted by
\cite[Liu \& Han (2014)]{liu14}. Attributing the jet power in the low accretion rate AGN to the BZ-jet,
this thus implies that the jet power has a flatter spectrum than the
accretion-dominated jet versus the accretion rate, as plotted in
Fig.~\ref{fig1} that the $P_{j}\propto (\lambda m)^{0.3}$ 
for the BH spin dominated jet in the Seyfert galaxies (the slope $\sim$0.3 from \cite[Liu \& Han
2014]{liu14}) has a flatter spectrum than the linear proportional relation for the disk
accretion-dominated jet power.

The linear proportionality of jet power versus
accretion rate is well constrained by the theory proposed by
\cite[Liu \& Han (2014)]{liu14}, and largely consistent with the
data as observed in \cite[van Velzen \& Falcke (2013)]{van13},
\cite[Fernandes et al. (2011)]{fer11}, and \cite[Willott et al.
(1999)]{wil99}, for radio loud quasars. However, in the lower
accretion regime, the power law slopes are obviously flatter ($\leq0.5$),  
see \cite[Liu \& Han (2014)]{liu14}, for radio galaxies and
Seyferts/LINERS. The flatter spectrum is most likely due to that the
BZ-jet power dominates over the accretion jet power in
the low accreting AGN. Of course, the BH spin must be high for
producing such jet power efficiently, and this implies that most
of radio-loud low-luminosity AGN may have high-spin BHs, and there are some
indications for this in \cite[Mart\'inez-Sansigre \& Rawlings
(2011)]{mar11} and \cite[Wu et al. (2011)]{wu11}. Contrarily, for the high
accretion quasars, their accretion jet power is dominating over their
BZ-jet power, so that we cannot identify whether their BH
is highly spinning or not from their jet luminosity only. 

This work is supported by the NSFC of
China (No.11273050).

\end{document}